\documentclass[prl,aps,twocolumn,floatfix,showpacs]{revtex4} 
\usepackage{psfig}
\usepackage{bm}
\usepackage{amsmath}

\newcommand{\be}{\begin{equation}}
\newcommand{\ee}{\end{equation}}
\newcommand{\bv}{{\bm v}}

\begin{document}

\title{Temperature probes in binary granular gases} 

\author{Alain Barrat$^1$, Vittorio Loreto$^2$ and Andrea
  Puglisi$^2$}

\affiliation{$^1$ Laboratoire de Physique Th{\'e}orique Unit{\'e}
Mixte de Recherche UMR 8627, B{\^a}timent 210, Universit{\'e} de
Paris-Sud, 91405 Orsay Cedex, France\\ $^2$''La Sapienza'' University,
Physics Department and INFM-SMC, Unit\`a di Roma1, P.le A. Moro 2,
00185 Rome, Italy}

\date{\today}

\begin{abstract}
We investigate the validity of Fluctuation-Dissipation (FD) relations for a
mixture of two granular gases with different physical properties
(restitution coefficients or masses) subject to stochastic driving. It
is well known that the partial granular temperatures $T_1$ and $T_2$
of the two components are different, i.e. energy equipartition is
broken. We observe, with numerical simulations of
inelastic hard disks in homogeneous and non-homogeneous situations,
that the classical equilibrium Green-Kubo relations are satisfied
separately for each component of the gas, the role of the equilibrium
temperature being played by the granular temperature of each
component. Of particular interest is 
the limit in which one of the two
components consists of only one particle, representing a
non-perturbing thermometer. In this case it turns out that such a
thermometer is measuring its own temperature and not that of the
surrounding granular media, which in general will be different.
\end{abstract} 

\maketitle 

\section{Introduction}

Granular gases~\cite{luding}, i.e. gases of particles interacting
through dissipative collisions, represent an important paradigm for
the study of non-equilibrium stationary states.  Due to the
dissipative nature of the interactions, granular gases have to be
considered as open systems and therefore concepts from equilibrium
thermodynamics cannot be applied, at least in a straightforward
way. However, by analogy with molecular gases, a ``granular
temperature'' $T_g$ can be defined in terms of the kinetic energy per
particle. If the system is driven by an external energy input a
stationary state is reached: $T_g$ fluctuates around a given value,
while a flux of energy from the driving source goes into the system
and is dissipated through inelastic collisions. Experiments, theory
and simulations have recently shown that in a mixture of different
grains, as soon as the inter-particle collisions dissipate energy, the
equipartition of energy among the mixture components is
lost~\cite{Duparcmeur,Losert,Menon,Wildman,Garzo,Clelland,
MontaneroHCS,Equipart,Puglisi2,Pagnani,Vibrated} even in the tracer
limit~\cite{Martin}. Although in sharp contrast with the behaviour of
molecular gases at equilibrium, this violation is not surprising
because in a generic open system the equipartition of energy is not
expected. However, it is natural to ask whether $T_g$ could have some
different ``equilibrium'' meaning, or if it is just a measure of velocity
fluctuations.  
A hallmark of equilibrium phenomena is the well-known
fluctuation dissipation theorem, relating the response of a system to
a perturbation to the corresponding correlation function measured in
the unperturbed system: the response of an observable $B$ at time $t$
to an impulsive perturbation $h_A$ at time $t=0$ can be obtained as
\begin{equation}
\frac{\delta \langle B \rangle}{\delta h_A}=
-\frac{1}{T}
\frac{\partial }{\partial t} \langle B(t) A(0) \rangle
\label{eq:fdt}
\end{equation}
where $A$ is the observable conjugated to $h_A$ and $T$ is the
equilibrium temperature of the system (brackets denote averaging
over thermal history). A recent investigation has
shown that, for non-equilibrium monodisperse driven granular gases,
this relation is still obeyed if the equilibrium temperature is
replaced by the granular temperature~\cite{FDT_granular}. 

A natural question then arises: what happens to the relation
(\ref{eq:fdt}) in a mixture which displays more than one granular
temperature ? In this Letter, we will address this question in the case
of a binary mixture subject to a homogeneous driving.  It turns out
that each species obeys an FDT relation with its own
granular temperature~\cite{deviations} as proportionality factor. This
result is particularly relevant to experiments where a tracer, immersed in the
granular gas and acting as a probe (and in particular as a
thermometer), may have different physical properties with respect to
the surrounding gas~\cite{Nature}. 
In equilibrium measurements, the temperature
does not depend on the thermometer. For granular materials, instead,
the thermometer measures its own temperature, in general different
from that of the surrounding gas.

\section{Model}

We consider a volume $V$ containing a mixture of $N=N_1+N_2$ inelastic
hard spheres (IHS) in dimension $d=2$, $N_1$ and $N_2$ being the
number of particles in component $1$ and $2$ of the mixture,
respectively. The spheres have diameters $\sigma$ (identical for the
two species) and masses $m_{s_i}$ (where $1 \leq i \leq N$ and $s_i$
is the species index, $1$ or $2$, of particle $i$).  In a collision
between spheres $i$ and $j$, characterized by the inelasticity
parameter called coefficient of normal restitution $\alpha_{s_is_j}$,
the pre-collisional velocity of particle $i$, ${\bm v}_i$, is
transformed into the post-collisional velocity ${\bm v}'_i$ such that
\begin{equation}
\bm{v}_i' \, =\,  \bm{v}_i - 
\frac{m_{s_j}}{m_{s_i}+m_{s_j}} (1+\alpha_{s_is_j})
(\widehat{\bm{\sigma}}\cdot \bm{v}_{ij})\widehat{\bm{\sigma}} 
\label{eq:coll1}
\end{equation}
where $\bv_{ij}=\bv_i-\bv_j$ and $\widehat{\bm\sigma}$ is the center
to center unit vector from particle $i$ to $j$
($\alpha_{s_is_j}=\alpha_{s_js_i}$ so that the total linear momentum $m_{s_i}
\bv_i+m_{s_j}\bv_j$ is conserved).
The granular temperature of species $s$ is given by its mean kinetic
energy $T_{s} = \langle m_{s} v_{s}^2\rangle/d$.

The loss of energy due to collisions can be compensated in various 
ways. In experiments the energy is typically supplied at the boundaries,
leading the system to a heterogeneous stationary 
state~\cite{Menon,Rouyer,Losert,Wildman}. 
In order to avoid the complication of strong temperature heterogeneities, we
will use a homogeneous driving in the form of a ``thermostat'': in this
mechanism (which recently has attracted the attention of many
theorists~\cite{Williams,Puglisi,Twan,Twanetal,Montanero,Cafiero,Moon}),
the particles are submitted, between collisions, to a random force in
the form of an uncorrelated white noise (e.g. Gaussian) with
the possible addition of a viscous term. The equation of motion for
a particle is then
\begin{equation}
m_i \frac{d {\bm v}_i}{dt} = {\bm F}_i + m_i {\bm R}_i - \zeta {\bm v}_i
\end{equation}
where ${\bm F}_i$ is the force due to inelastic collisions, $\zeta$ is
the viscosity coefficient and $\langle R_{i\alpha}(t) R_{j\beta}(t')
\rangle = \xi_0^2 \delta_{ij}\delta_{\alpha\beta} \delta(t-t')$, where
Greek indexes refer to Cartesian coordinates \cite{deterministic}.

At the level of Boltzmann kinetic equation, the temperature ratio of a
binary granular mixture subject to stochastic driving of the form
given above has been obtained in~\cite{Equipart} for the case
$\zeta=0$ and in~\cite{Pagnani} for $\zeta \neq 0$.

\section{Methods}

We have used two different simulation methods: the Direct Simulation
Monte Carlo (DSMC)~\cite{Bird,mueller} which neglects pre-collisional
correlations and therefore enforces the molecular chaos hypothesis
(factorization of the two-particles distribution functions) and
Molecular Dynamics (MD) simulations. The DSMC algorithm can be used to
simulate the {\em homogeneous} Enskog-Boltzmann equation, i.e. any
particle can collide with any other in the whole volume $V$, or to
simulate an a priori {\em non-homogeneous} system, by dividing the space into
small cells (of linear size smaller than the mean free path) and
enforcing molecular chaos in every single cell. On the other hand,
molecular dynamics (MD) simulations integrate the equations of motion
of the IHS: in this case we consider $N$ hard spheres in a square box
of linear size $L=\sqrt{V}$, with periodic boundary conditions, random
initial velocities, and we use an event-driven algorithm to study
their dynamics. All the investigation methods used have shown similar
results.  We have used sizes $N=500$ to $N=5000$, and averaged over up
to $10000$ realizations of the dynamics.


Our analysis relies on two sets of independent measurements, i.e. two
choices of the pair response-correlation to be measured. The first one
consists in a measure of mobility and diffusivity. The mean-square
displacement (MSD) for each species,
\begin{equation}
B_{1(2)}(t,t') = \frac{1}{d N_{1(2)}}\sum_{j=1}^{N_{1(2)}} 
\langle |{\bm r_j}(t) - {\bm r_j}(t')|^2 \rangle  \ ,
\end{equation}
behaves as $\sim 4 D_{1(2)}(t-t')$ for large time differences. The
mobility of a tracer particle is measured by applying a constant and
small~\cite{linearresponse} drag force ${\bm \xi}=\xi {\bm e_x}$ to a
given particle, labeled $0$, for times $t > t'$.  Due to the viscosity
induced by the collisions with other particles, the perturbed particle
will reach at large times a constant velocity $\mu$, related to the response by
\begin{equation}
\chi_{1,2} (t,t') = \frac{1}{\xi} \langle ({\bm r_0 (t)} - {\bm r_0
(t')})\cdot {\bm e_x} \rangle \approx \mu t \ \ \mbox{at large\ t}.
\end{equation}
By successively using as test particle one particle of each species,
one obtains the two responses $\chi_1$ and $\chi_2$, and thus the
mobilities $\mu_1$ and $\mu_2$. Two Einstein relations ($\mu_i= 2 D_i
/T_i $) can then be checked, e.g. by plotting $\chi_i$ vs. $B_i$.

Another totally independent way of checking FD relations in granular
gases has also been used in~\cite{FDT_granular}: once a steady-state
has been reached, the system is perturbed impulsively at a given time
$t_0$ by a non-conservative force applied (non-uniformly) on every
particle (we will take $t_0=0$ without loss of
generality). The response is then monitored at later times. The force
acting on particle $i$ is
\begin{equation} \label{forcing}
\mathbf{F}(\mathbf{r}_i,t)=\gamma_i \boldsymbol{\xi}(\mathbf{r}_i,t)
\end{equation}
with the properties $\boldsymbol{\nabla} \times \boldsymbol{\xi} \neq
0$, $\boldsymbol{\nabla} \cdot \boldsymbol{\xi} = 0$, where $\gamma_i$
is a particle dependent variable with randomly assigned $\pm 1$
values. A simple case is realized by a transverse perturbation ${\bm
\xi}(\mathbf{r},t)= (0, \xi \cos (k_x x) \delta(t-t_0)
)$~\cite{linearresponse}, where $k_x$ is compatible with the periodic
boundary conditions, i.e. $k_x=2 \pi n_k /L_x$ with $n_k$ integer and
$L_x$ the linear horizontal box size.  The staggered response function
(i.e. the current induced at $t$ by the perturbation at $t_0$), and the
conjugated correlation,
\begin{eqnarray} 
R(t,t_0) & = & \frac{1}{\xi}\langle \sum_i \gamma_i \dot{y}_i(t)
\cos(k_x x_i(t)) \rangle \ ,\nonumber \\ 
C(t,t_0) & = &\langle \sum_i
\dot{y}_i(t) \dot{y}_i(t_0) \cos \{k_x[x_i(t)-x_i(t_0)] \} \rangle
\nonumber
\end{eqnarray}
are related, {\em at equilibrium}, to the correlation by the FD
relation $R(t,t_0) = \frac{\beta}{2} C(t,t_0)$, $T=1/\beta$ being the
equilibrium temperature.

Ref.~\cite{FDT_granular} has shown the validity of this relation
in the context of a monodisperse granular media heated by a thermal
bath with temperature $T_b$, reaching in this way a stationary state
with granular temperature $T_g<T_b$. In this case the FD relation holds 
by replacing $T$ with $T_g$. For a binary
mixture, two sets $(C_i, R_i)$ ($i=1,2$) of correlation and response
are measured separately, thus obtaining two plots $C_i$ vs. $R_i$.

There are different reasons to test various pairs of response and correlation
functions. First, at equilibrium FDT links any couple of conjugated
correlation and response function with the {\em same} temperature. Out of
equilibrium, it is possible, a priori, that FDT could be valid for some
observables and not for others. Moreover, mean square displacement and
mobility correspond to the exploration of the long time regime, while $C_i$
and $R_i$ decay quickly (a few collisions per particle) and therefore yield
the short time behaviour. FDT could be valid in some time regimes and not in
others.  It must also be stressed that the measure of diffusion and mobility
requires large times, so that averaging over many realizations becomes
computationally very demanding and less precision is numerically available. In
MD simulations in particular we have measured only the relation between $C_i$
and $R_i$.

\section{Results}

In all the simulations performed (MD and DMSC) the values of the temperature
ratios $T_1/T_2$ obtained are in good agreement with~\cite{Equipart}. This is
not surprising for homogeneous DSMC, while it is less obvious in MD (the
agreement is mostly due to the low packing fraction $n$ used).

\subsection{Fluctuation-Dissipation ratio for the different components} 

\begin{figure}[htb]
\centerline{
        \psfig{figure=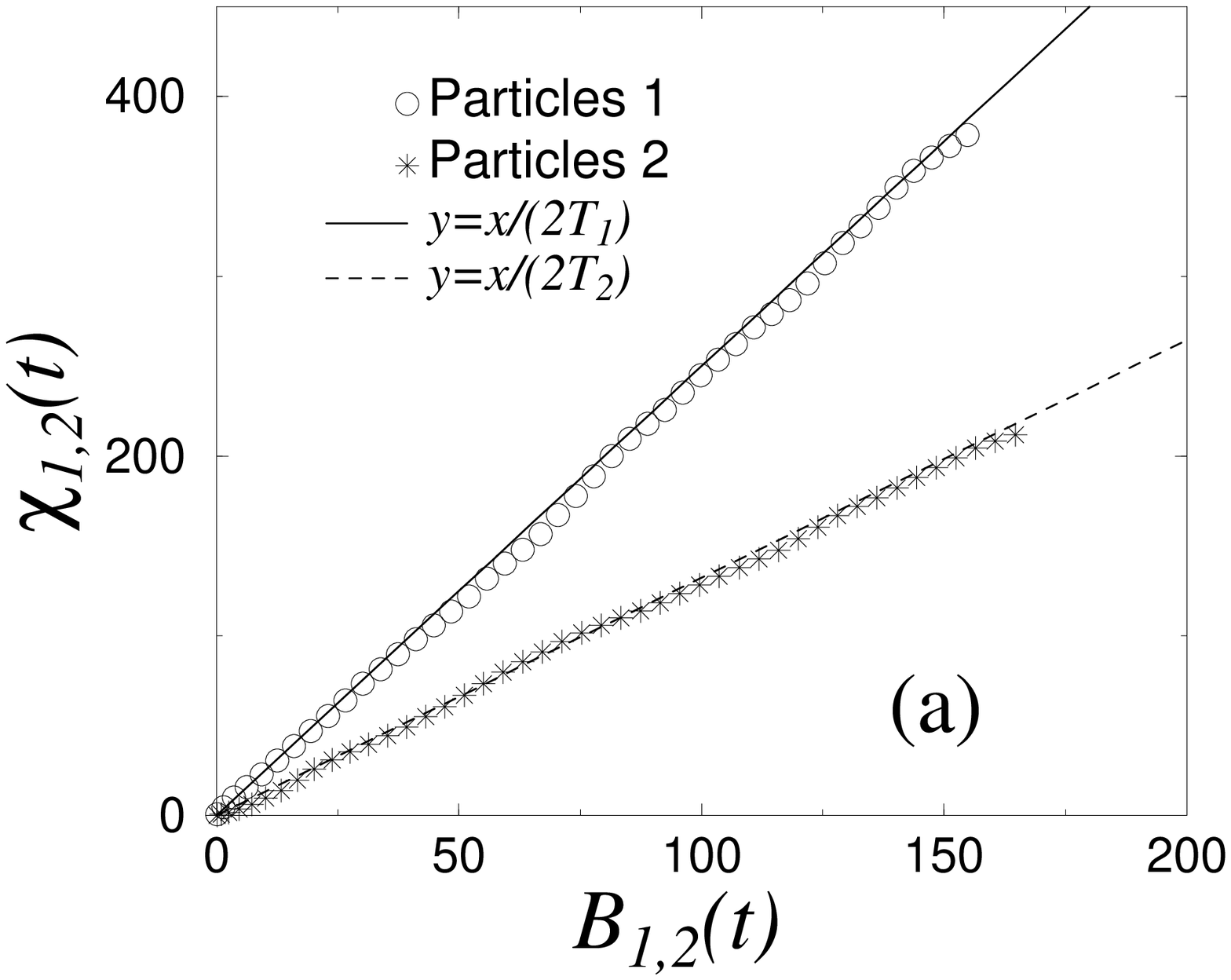,width=4cm,angle=0}
        \psfig{figure=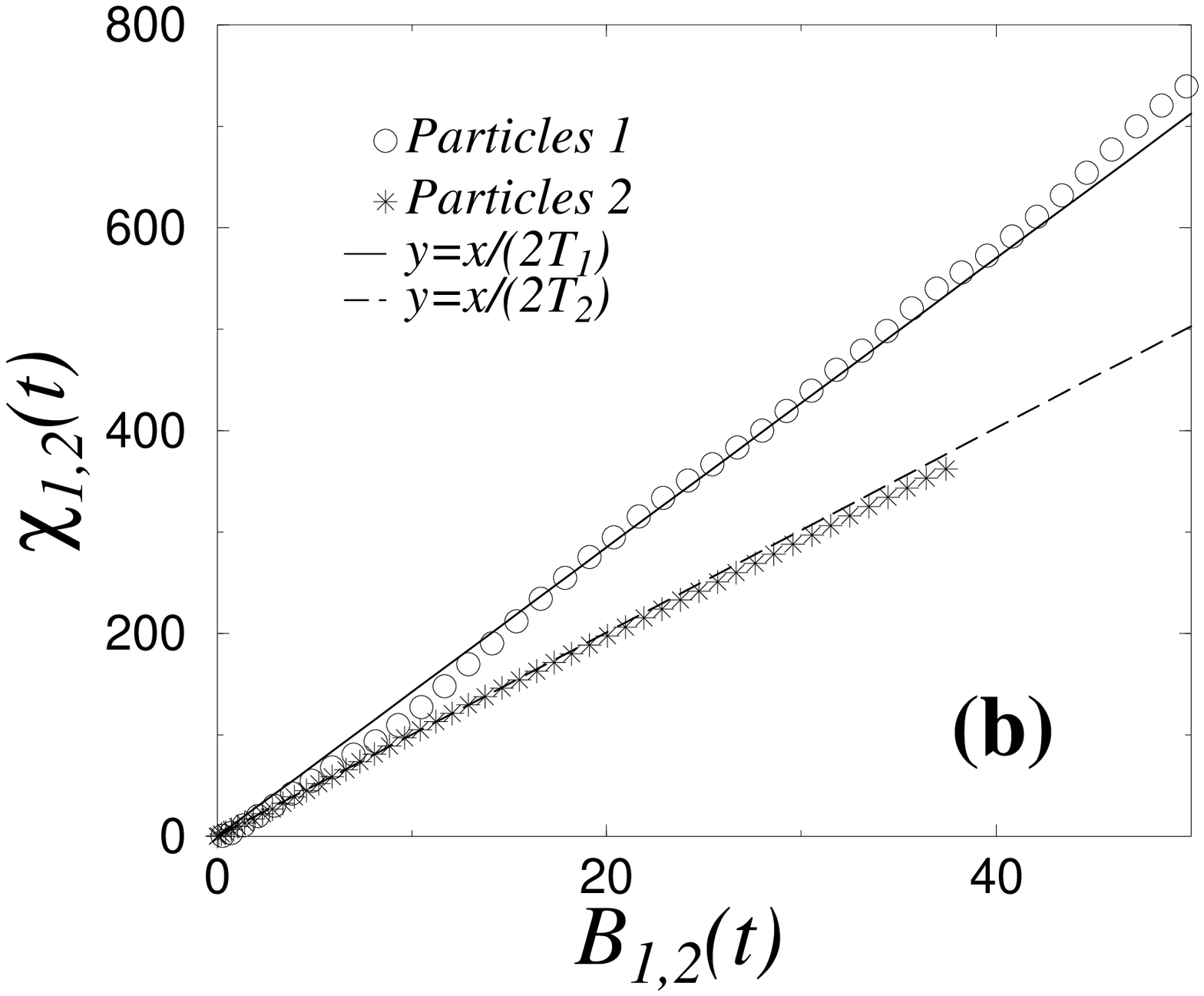,width=4cm,angle=0}
}
\caption{Random force ($\zeta=0$), 
homogeneous DSMC: mobilities $\chi_{1,2}$ vs. MSD
$B_{1,2}$; (a): $\alpha_{11}=0.3$, $\alpha_{12}=0.5$,
$\alpha_{22}=0.7$, $m_2=3m_1$, $T_1\approx 0.2$, $T_2\approx 0.38$;
(b): $\alpha_{11}=\alpha_{12}=\alpha_{22}=0.9$, $m_2=5m_1$,
$T_1\approx 0.035$, $T_2\approx 0.05$.  Symbols are numerical data,
lines have slope $1/(2T_1)$ and $1/(2T_2)$.  }
\label{fig:hom_kicks}
\end{figure}

\begin{figure}[htb]
\centerline{
        \psfig{figure=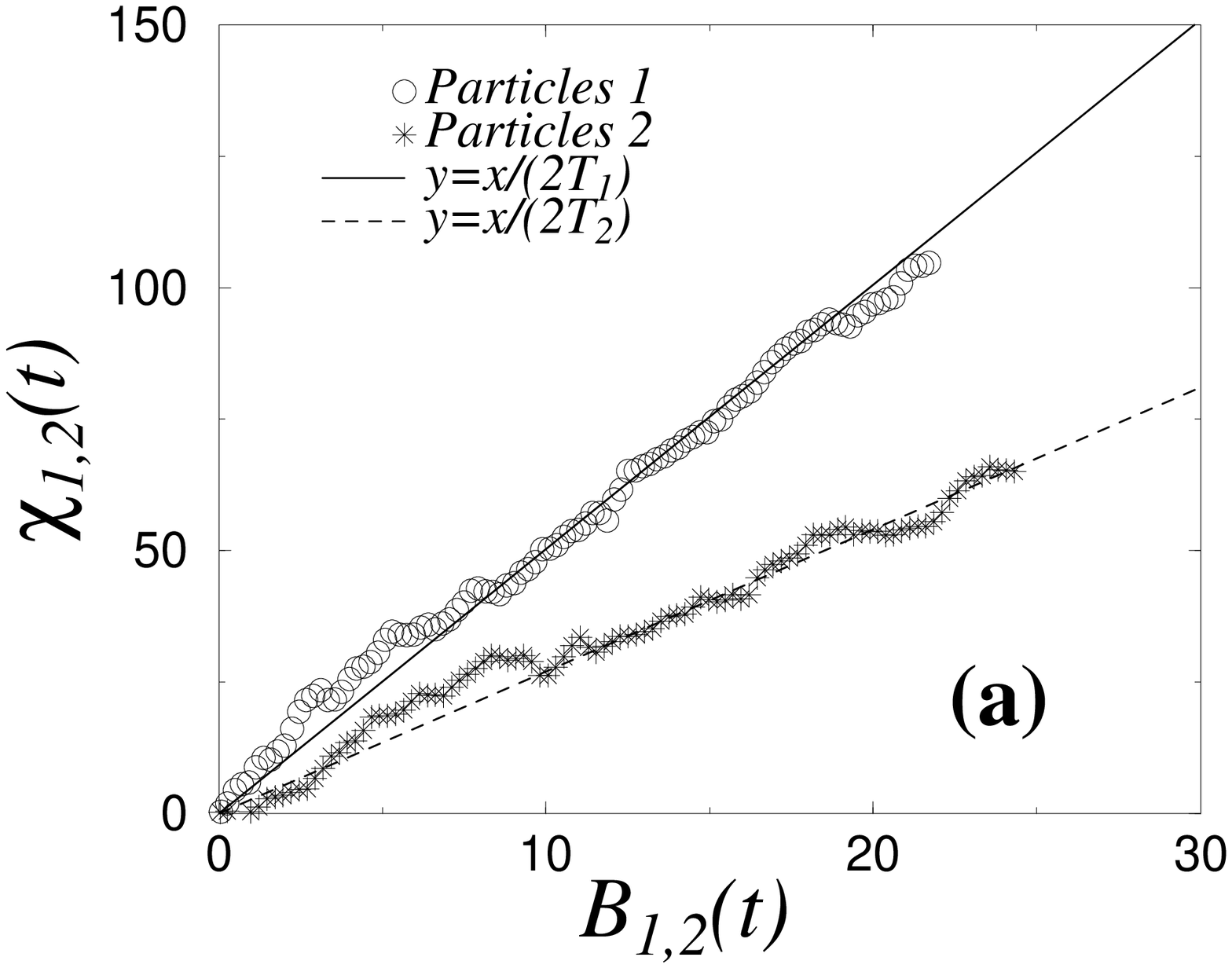,width=4cm,angle=0}
        \psfig{figure=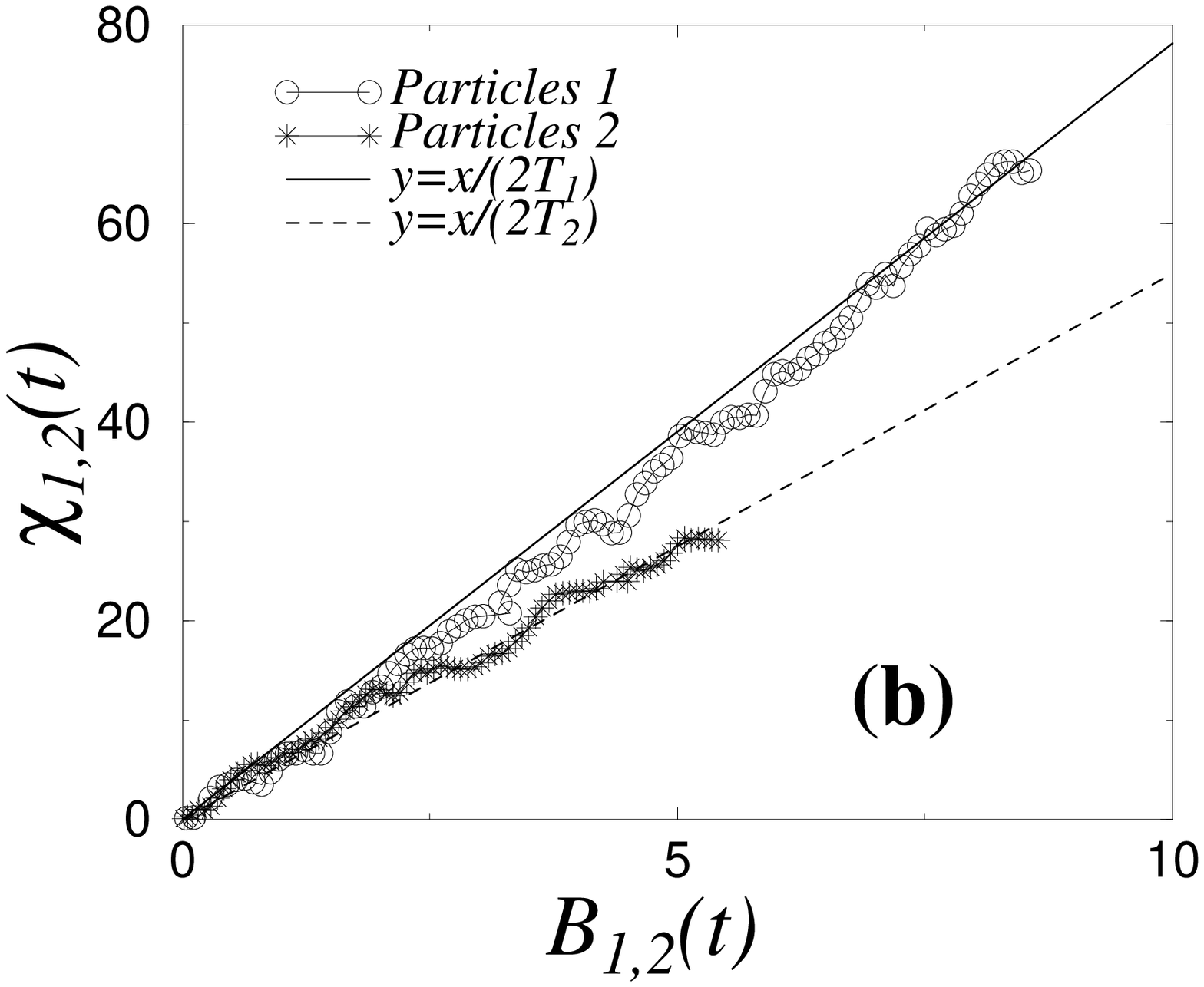,width=4cm,angle=0}
}
\caption{Random force ($\zeta=0$), 
DSMC with cells: mobilities $\chi_{1,2}$ vs. MSD
$B_{1,2}$. (a): $\alpha_{11}=0.3$, $\alpha_{12}=0.5$,
$\alpha_{22}=0.7$, $m_2=3m_1$, $T_1\approx 0.1$, $T_2\approx 0.185$;
(b): $\alpha_{11}=\alpha_{12}=\alpha_{22}=0.9$, $m_2=5m_1$,
$T_1\approx 0.064$, $T_2\approx 0.09$.  Symbols are numerical data,
lines have slope $1/(2T_1)$ and $1/(2T_2)$.}
\label{fig:cells_kicks}
\end{figure}

The measures of $B_i(t)$ and $\chi_i(t)$ allow us to check
the linearity with $t$ at large times:  $B_i(t) \approx 4 D_i t$ 
and $\chi_i(t) \approx \mu_i t$. Moreover,
figures~\ref{fig:hom_kicks} and~\ref{fig:cells_kicks} clearly show that the
Einstein relations are obeyed separately by the two components of the mixture,
each with its own temperature: to the numerical accuracy one obtains 
$\chi_i(t) = \frac{1}{2T_i}B_i(t)$. Note that mobilities and diffusion
coefficients of each species are different from their value in a monodisperse
case.  We have considered various values of inelasticities, mass ratios,
number density ratios, and kinds of thermostat, obtaining that this result is
robust with respect to all these variations. We also note that a recent
study~\cite{Dufty} has shown that the Einstein relation is not valid in its 
usual form in the case of an impurity immersed in an homogeneously cooling 
granular: the main reason is that extra terms arise due to the evolution
of the granular temperature with time. Since 
we are here concerned with {\em steady-states}, 
these extra terms are not present in the case studied. The other
source of deviations, namely the deviation from Gibbs state,
leads to small deviations~\cite{Dufty} that could be difficult to
detect~\cite{deviations}.

\begin{figure}[htb]
\centerline{
        \psfig{figure=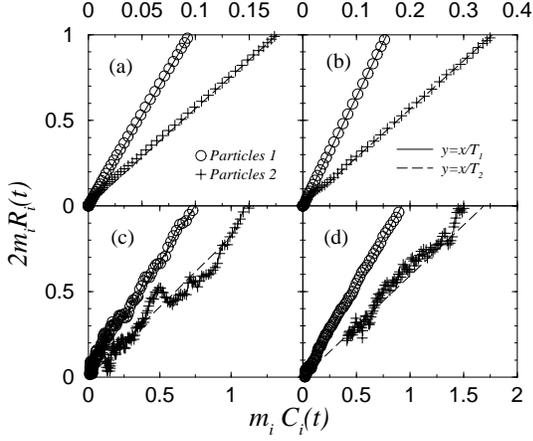,width=7cm,angle=0}
}
\caption{Random force (i.e. $\zeta=0$), 
DSMC with cells, and MD: $2 m_i R_i(t)$ vs. $m_i
C_i(t)$ ($i=1,2$). Here $n_k=4$. $C_i(0)=T_i/m_i$,
$R_i(0)=1/(2m_i)$. In all cases the circles correspond to the FD plot
for the type 1 particles, the pluses to the FD plot for type 2
particles. Straight lines have slope $1/T_1$, dashed lines have slope
$1/T_2$.  {\bf Top}: DSMC simulations. (a): $\alpha_{11}=0.3$,
$\alpha_{12}=0.5$, $\alpha_{22}=0.7$, $m_2=3m_1$, $T_1\approx 0.09$,
$T_2 \approx 0.18$; (b): $\alpha_{11}=\alpha_{12}=\alpha_{22}=0.7$,
$m_2=5m_1$, $T_1\approx 0.16$, $T_2 \approx 0.35$; {\bf Bottom}: MD
simulations, with $n=0.1$.  (c): $N_1=N_2$, $\alpha_{11}=0.7$,
$\alpha_{12}=0.8$, $\alpha_{22}=0.9$, $m_2=3m_1$, $T_1\approx 0.75$,
$T_2 \approx 1.22$; (d): $N_1=9N_2$;
$\alpha_{11}=\alpha_{12}=\alpha_{22}=0.9$, $m_2=5m_1$, $T_1 \approx
0.93$, $T_2\approx 1.69$.}
\label{fig:RC_DSMC_MD}
\end{figure}

We now turn to the measure of $R_i(t)$ and $C_i(t)$. Technical details
of the numerical procedure to perform this measure are given
in~\cite{FDT_granular}.  From the definitions of ${\bm \xi}$, $C_i$
and $R_i$, it is clear that $R_i(0)=1/(2m_i)$, and $C_i(0)= \langle
v_i^2 \rangle = T_i/m_i$. On the other hand, $\lim_{t \to
\infty}R_i(t)=\lim_{t \to \infty}C_i(t)=0$.  Thus we have plotted in
figure~\ref{fig:RC_DSMC_MD} the functions $2m_iR_i(t)$
vs. $m_iC_i(t)$. The FD relation $R_i(t) = \frac{1}{2T}
C_i(t)$ is verified replacing $T$ by the partial granular
temperature $T_1$ and $T_2$ of each component. The same result holds
for different $n_k$ ($n_k \neq 0$ in order to satisfy the properties
of the perturbing force~(\ref{forcing})).

The MD simulations (whose results are reported in
figure~\ref{fig:RC_DSMC_MD}) are more realistic since they include
excluded volume effects and collision-induced correlations (which
break the Molecular Chaos hypothesis, see for
example~\cite{Twanetal}). However, at not too high packing fractions,
we still observe the same results for FD relations. Larger packing
fractions lead to strong heterogeneities in both density and granular
temperature, giving rise to deviations from FD~\cite{deviations}.

\subsection{Fluctuation-Dissipation ratio for the whole system}

An interesting question concerns what happens to the Fluctuation-Dissipation
ratio when measured for the whole system and not separately for the different
components of the mixture. In other words one could ask whether one can define
an effective temperature for the whole system and what is the relation of this
temperature with the temperatures defined above for the two components of
the mixture, or with the global temperature
$T=x_1 T_1 + x_2 T_2$ (where $x_{1,2}=\frac{N_{1,2}}{N}$).

A global measure would give, for the mean-square displacement 
\begin{equation} 
B(t) = x_1 B_1(t)+ x_2 B_2(t)\ ,
\end{equation}
and for the response
\begin{equation} 
\chi(t) =x_1 \chi_1(t)+ x_2 \chi_2(t)\ .
\end{equation}

\begin{figure}[htb]
\centerline{
        \psfig{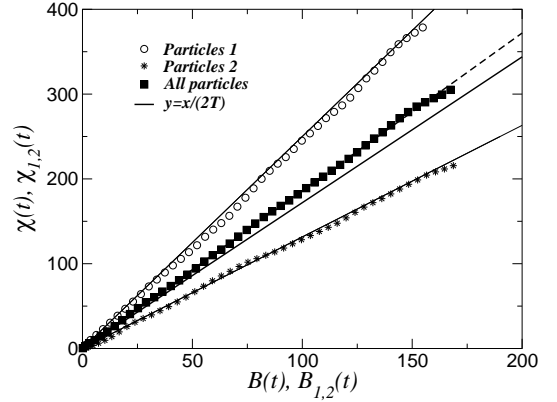}
}
\caption{Homogeneous DSMC, random force,
  $\alpha_{11}=0.3, \alpha_{12}=0.5, \alpha_{22}=0.7$, $m_2=3m_1$, $N_1=N_2$.
  The filled squares correspond to the numerical data of $\chi(t)$ vs. $B(t)$.
  Circles and stars correspond respectively to particles of type $1$ and $2$,
  and the thin lines have slopes $1/(2T_1)$ and $1/(2T_2)$.  The thick
  continuous line has slope $1/(2T)$ while the dashed line has slope
  $\mu/(4D)$ with $2D/\mu$ given by eq.~\ref{eq:global}.
Here $T \approx 0.29$ while $2D/\mu \approx 0.27$.}
\label{fig_einstein_global}
\end{figure}

Using the previously checked result that $\chi_i(t) \approx \mu_i t$ and
$B_i(t) \approx 4D_i t$, one obtains that $B(t) \approx 4D t$ and $\chi(t)
\approx \mu t$, with $D= x_1 D_1 + x_2 D_2$ and $\mu= x_1 \mu_1 + x_2 \mu_2$.
The Einstein relation for the global case thus reads:
\begin{equation} 
\frac{2 D}{\mu} = 
\frac{(x_1 D_1 + x_2 D_2) T_1 T_2}{x_1 D_1 T_2+ x_2 D_2 T_1},
\label{eq:global}
\end{equation}
It is clear that the ratio $\frac{2 D}{\mu}$ corresponds to the global
granular temperature only when $T_1= T_2$, i.e. when equipartition is
satisfied. In figure~\ref{fig_einstein_global} we present 
evidences supporting this view.

\vskip .5cm

\begin{figure}[htb]
\centerline{
        \psfig{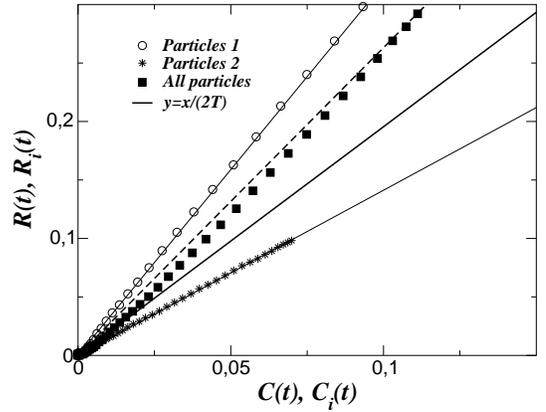}
}
\caption{DSMC with cells, random force,
  $\alpha_{11}=\alpha_{12}=\alpha_{22}=0.7$, $m_2=5m_1$, $N_1=N_2$.  The
  filled squares correspond to the numerical data of $R(t)$ vs. $C(t)$.
  Circles and stars correspond respectively to $R_1(t)$ vs. $C_1(t)$ and
  $R_2(t)$ vs. $C_2(t)$.  Thin lines have slopes $1/(2T_1)$ and $1/(2T_2)$.
  The thick continuous line has slope $1/(2T)$; the dashed line is a guide to
  the eye showing that $R(t)$ vs. $C(t)$ is not a straight line.}
\label{fig_fdr_gobal}
\end{figure}

Another way to consider the problem is to look at the ratio between
$R(t)=x_1 R_1(t)+ x_2 R_2(t)$ and $C(t)=x_1 C_1(t)+ x_2
C_2(t)$. In this case the impossibility to define an effective temperature
reflects itself in the non-constant ratio between $C(t)$ and $R(t)$, 
as we show in figure~\ref{fig_fdr_gobal}. 
\vskip .5cm

It should be remarked how these properties of the global
Fluctuation-Dissipation ratio allow us to make the following prediction.
Suppose to perform a global measurement of the Fluctuation-Dissipation ratio
on a system of unknown composition. A global measurement on a monodisperse
system would yield a well defined FD ratio and a well defined temperature
equal to the granular temperature, independently of the observable used (see
\cite{FDT_granular}). On the other hand a polydisperse system would feature a
typically non-constant FD ratio, function also of the observable (unless the
system is completely elastic).

\subsection{The single tracer case}

\begin{figure}[tb]
\centerline{ \psfig{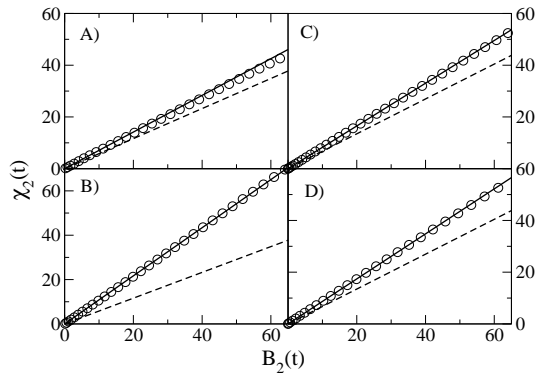} }
\caption{ Circles: Mobility vs. Diffusion of a single particle of mass
$m_{tracer}$ in contact with $N=500$ particles of mass $m$, immersed
in a heat bath (i.e. random kicks plus viscosity).  We use the
following conventions: $\alpha_{tracer}=\alpha_{12}$ and
$\alpha=\alpha_{11}$. Only in case A) the tracer is also in contact
with the external driving heat bath.  {\bf A)} $m_{tracer}=m$,
$\alpha=0.9$, $\alpha_{tracer}=0.4$, $T_g=0.86$, $T_g^{tracer}=0.70$;
{\bf B)} $m_{tracer}=m$, $\alpha=0.9$, $\alpha_{tracer}=0.4$,
$T_g=0.86$, $T_g^{tracer}=0.46$; {\bf C)} $m_{tracer}=7m$,
$\alpha=\alpha_{tracer}=0.7$, $T_g=0.74$, $T_g^{tracer}=0.60$; {\bf
D)} $m_{tracer}=4m$, $\alpha=\alpha_{tracer}=0.7$, $T_g=0.74$,
$T_g^{tracer}=0.57$.  The solid line has slope $1/(2T_g^{tracer})$, the
dashed line has slope $1/(2T_g)$.  }
\label{fig:tracer}
\end{figure}

Finally we investigate the special case $N_2=1$, i.e. the case of a single
tracer, immersed in a granular gas of $N=N_1$ different particles, acting as a
non-perturbing thermometer. Figure~\ref{fig:tracer} reports the corresponding
results. It turns out that, due to the inelasticity of
collisions~\cite{Martin}, the tracer reaches a granular temperature different
from that of the surrounding particles. Measuring a FD relation thus yields
the granular temperature of the tracer but not that of the surrounding
granular gas. The tracer is thus sensing a temperature whose value is the
outcome of the complex interaction between the tracer itself and the granular
gas. It is worth stressing how this happens {\em even if the tracer almost
  does not perturb the granular}. This leads to the non-conventional result
that the ``temperature'' measured as a fluctuation-dissipation
ratio~\cite{deviations}, in a driven granular gas, depends on the thermometer.

\section{Conclusions}

In this paper, we have shown by numerical simulations that, in a binary
granular gas, each component of the mixture obeys a FD relation with its own
granular temperature, while a global measurement provides a suitable
definition of temperature only when equipartition is satisfied. These FD
relations can be measured by different correlation-response pairs. In
particular, the case of a tracer, which can act as a ``thermometer'' since it
does not perturb the granular, has been investigated, and leads to the unusual
conclusion that the measure of the temperature through FD relations would
depend on the interaction between the tracer and the granular gas. Notice that
theoretical studies (see e.g.~\cite{Equipart}) may then allow for an estimate
of the gas temperature, knowing the temperature measured by the tracer.
Further investigations are needed to explore the scenarios bringing to
violations of the FD relations~\cite{deviations}. This point becomes
particularly important when, with more realistic energy injection through
boundaries, we could expect heterogeneities giving rise to position-dependent
temperatures. It would be interesting to investigate how a tracer particle
would sample this inhomogeneous gas and ask about the meaning of the
corresponding measured temperature.

{\large Acknowledgments} We are grateful to A. Baldassarri, H. J. Herrmann
and E. Trizac for many enlightening discussions.


\begin{thebibliography}{99}

\bibitem{luding} T.~P\"oschel and  S.~Luding (Eds.),  {\em Granular Gases},
  Springer, Berlin (2001).

\bibitem{Duparcmeur}
Y. Limon Duparcmeur, Th\`ese de l'universit\'e de Rennes I (1996).

\bibitem{Losert}
W. Losert, D.G.W. Cooper, J. Delour, A. Kudrolli and J.P. Gollub,
Chaos {\bf 9}, 682 (1999).

\bibitem{Menon}
K. Feitosa and N. Menon, 
Phys. Rev. Lett. {\bf 88}, 198301 (2002).

\bibitem{Wildman}
R.D. Wildman and D.J. Parker, 
Phys. Rev. Lett. {\bf 88}, 064301 (2002).

\bibitem{Garzo}
V. Garz\'o and J. Dufty,
Phys. Rev. E {\bf 60} 5706 (1999).

\bibitem{Equipart}
A. Barrat and E. Trizac, 
Gran. Matter {\bf 4}, 57 (2002).

\bibitem{MontaneroHCS}
J. M. Montanero and V. Garz\'o, 
Gran. Matter {\bf 4}, 17 (2002).

\bibitem{Clelland}
R. Clelland and C. M. Hrenya, 
Phys. Rev. E {\bf 65}, 031301 (2002).

\bibitem{Puglisi2}
U. Marini Bettolo Marconi and A. Puglisi,
Phys. Rev. E {\bf 65}, 051305 (2002)
and Phys. Rev. E {\bf 66}, 011301 (2002).

\bibitem{Pagnani}
R. Pagnani, U.M. Bettolo Marconi and A. Puglisi,
Phys. Rev. E {\bf 66}, 051304 (2002).

\bibitem{Vibrated}
A. Barrat and E. Trizac, 
Phys. Rev. E {\bf 66}, 051303 (2002). 

\bibitem{Martin}
Ph. A. Martin and J. Piasecki, Europhys. Lett. {\bf 46}, 613 (1999).

\bibitem{FDT_granular}
A. Puglisi, A. Baldassarri, and V. Loreto 
Phys. Rev. E {\bf 66}, 061305 (2002).


\bibitem{deviations} There exist in fact small, but systematic,
deviations from FD relations even at small inelasticities. These
deviations appear even at the homogeneous Boltzmann level
(investigated by DSMC), where they are linked to the non-Gaussian
behaviour of the velocity distribution due to the
departure from equilibrium~\cite{Dufty}. The departure from
(\ref{eq:fdt}) will be studied in a separate publication.

\bibitem{Dufty}
J. Dufty and V. Garz\`o, J. Stat. Phys. {\bf 105}, 723 (2001).

\bibitem{Nature}
G. D'Anna, P. Mayor, A. Barrat, V. Loreto and F. Nori,
Nature {\bf 424}, 909 (2003).

\bibitem{Rouyer}
F. Rouyer and N. Menon, Phys. Rev. Lett. {\bf 85}, 3676 (2000).

\bibitem{Williams}
D.R.M. Williams and F.C. MacKintosh, Phys. Rev E {\bf 54}, R9 (1996).

\bibitem{Puglisi}
A. Puglisi, V. Loreto, U. Marini Bettolo Marconi, A. Petri and
A. Vulpiani, Phys. Rev. Lett {\bf 81}, 3848 (1998); A. Puglisi,
V. Loreto, U. Marini Bettolo Marconi and A. Vulpiani, Phys. Rev. E
{\bf 59}, 5582 (1999).

\bibitem{Twan}
T.P.C. van Noije and M.H. Ernst,  Gran. Matter {\bf 1}, 57 (1998).

\bibitem{Twanetal}
T.P.C. van Noije, M.H. Ernst, E. Trizac and I. Pagonabarraga,
Phys. Rev. E {\bf 59}, 4326 (1999);
I. Pagonabarraga, E. Trizac, T.P.C. van Noije, and M.H. Ernst,
Phys. Rev. E {\bf 65}, 011303 (2002).

\bibitem{Montanero}
J.M. Montanero and A. Santos, Granular Matter {\bf 2}, 53 (2000).

\bibitem{Cafiero}
R. Cafiero, S. Luding and H.J. Herrmann,
Phys. Rev. Lett. {\bf 84}, 6014 (2000).

\bibitem{Moon}
S.J. Moon, M.D. Shattuck and J.B. Swift, Phys. Rev. E {\bf 64}, 031303 (2001).

\bibitem{deterministic} It is also possible to use a deterministic
thermostat. See e.g.: D. J. Evans, G. P. Morriss, {\em
Statistical Mechanics of Nonequilibrium Liquids}, (Academic Press,
London, 1990).

\bibitem{linearresponse}
We have checked the linearity of
the response by changing the amplitude of the perturbation.

\bibitem{Bird} G. Bird, {\em Molecular Gas Dynamics}, (Oxford
University Press, New York, 1976) and {\em Molecular Gas Dynamics and
the Direct Simulation of Gas flows}, (Clarendon Press, Oxford, 1994).

\bibitem{mueller} M.S. M\"uller, {\em Fast algorithms for the simulation
of granukar particles}, Ph.D. Thesis, Univ. of Stuttgart (2001); M.S. M\"uller
and H.J. Herrmann, in {\em Physics of Granular Media}, H.J. Herrmann,
J.-P. Hovi and S. Luding, NATO ASI Series, Kluwer Academic Publisher,
Dordrecht (1998).

\end{thebibliography}
\end{document}